\documentclass[a4paper,11pt]{article}
\usepackage{jheppub} 

\usepackage[usenames,dvipsnames]{xcolor}
\usepackage{amsmath,amssymb,bbold,epsfig}
\usepackage{amsfonts}
\usepackage{slashed}
\usepackage{dsfont}
\usepackage{graphicx,color}
\usepackage{bbm}
\usepackage{bm}
\usepackage[normalem]{ulem}
\definecolor{lightgray}{gray}{0.85}

\vspace*{-1.5cm}
\title{
Neutrino oscillations in matter: from microscopic to  
macroscopic description}

\author{Evgeny~Akhmedov}

\affiliation{
Max-Planck-Institut f\"ur Kernphysik, Saupfercheckweg 1, \\69117 Heidelberg,
Germany
}

\emailAdd{akhmedov@mpi-hd.mpg.de}

\abstract{Neutrino flavour transmutations in nonuniform matter are described 
by a Schr\"{o}dinger-like evolution equation with coordinate-dependent 
potential. In all the  derivations of this equation it is assumed that the 
potential, which is due to coherent forward scattering of neutrinos on 
matter constituents, is a continuous function of 
coordinate that changes slowly over the distances 
of the order of the neutrino de Broglie wavelength. 
This tacitly assumes that some averaging of the microscopic potential 
(which takes into account the discrete nature of the scatterers) 
has been performed. 
The averaging, however, must be applied to the microscopic evolution equation 
as a whole and not just to the potential. Such an averaging  
has never been explicitly carried out. We fill this gap by considering 
the transition from the microscopic to macroscopic neutrino evolution 
equation through a proper averaging procedure. We discuss 
some subtleties related to 
this procedure and establish the applicability domain of the 
standard macroscopic evolution equation.
This, in particular, allows us to answer the question of when neutrino 
propagation in rarefied media (such as e.g.\ 
low-density gases or interstellar or intergalactic media) can be considered 
within the standard theory of neutrino flavour evolution in matter.  
}
\begin{document}
\maketitle
\flushbottom

\newcommand{\be}{\begin{equation}}
\newcommand{\ee}{\end{equation}}
\newcommand{\bea}{\begin{eqnarray}}
\newcommand{\eea}{\end{eqnarray}}

\section{\label{sec:intro}Introduction}

Neutrino flavour transformations in nonuniform matter are 
described by a Schr\"{o}dinger-like evolution equation 
with coordinate-dependent potential. It had been first suggested by 
Wolfenstein \cite{Wolf} basing on heuristic 
considerations and was subsequently derived more rigorously within the 
relativistic quantum mechanics and quantum field theory frameworks 
\cite{Halprin,mannheim,changzia,sawyer,grimus,cardchung,AW}.  
This evolution equation 
has been employed in virtually all studies of the neutrino 
flavour transition effects in nonuniform matter, including the explorations 
of the Mikheyev-Smirnov-Wolfenstein effect \cite{Wolf,MS} and 
of the 
parametric resonance of neutrino oscillations 
\cite{param1,param2,param3} (see ref.~\cite{BS} for a review). 
 
In all the  derivations of the neutrino evolution equation it 
is assumed that the neutrino potential, which is 
due to coherent forward scattering of neutrinos on matter constituents, 
is a continuous function of coordinate that changes slowly over 
the distances of the order of the neutrino de Broglie wavelength 
$\lambda_D=1/p$. 
This means that some averaging of the microscopic potential 
(which takes into account the discrete nature of the scatterers) 
is tacitly assumed. 

Indeed, even for neutrinos of energy as small as $E\sim 1$ MeV 
(which is close to the lower end of the spectra of detectable 
solar neutrinos and reactor antineutrinos) 
the de Broglie wavelength is on the order of 
$10^{-11}$ cm, which is much smaller than the 
interatomic distance $\sim 10^{-9}$--$10^{-8}$ cm.
Thus, the average number of scatterers inside a volume 
of linear size $\sim\lambda_D$ is much less 
than one, and the matter-induced neutrino potential cannot even approximately 
be considered as a smooth function on such length scales.  
This means that some coarse-graining (averaging) must be performed 
to justify the standard neutrino evolution equation. 
It is important to note that such an averaging must be applied to 
the microscopic evolution equation as a whole and not just to the potential. 
No such averaging has been explicitly carried out so far.   

In the present paper we fill this gap by considering the transition from 
the microscopic to macroscopic neutrino evolution equation through a 
proper averaging procedure. We perform a coarse graining -- a 
coordinate-space averaging over macroscopic volumes $v_0$ that contain 
large numbers of particles of the medium and at the same time are small 
enough, so that the macroscopic characteristics of the medium (such as 
density) are nearly constant within $v_0$.%
\footnote{In some situations also the averaging over the velocities 
and spins of the particles of the medium has to be done. This is discussed in 
section~\ref{sec:spatial}.}
Such an averaging is actually necessary 
because of the very large number of the scatterers, which makes a microscopic 
description of neutrino flavour evolution practically impossible; 
it is also sufficient, as we only need  the coarse-grained neutrino wave 
functions to predict the outcomes of neutrino detection experiments.

The averaging procedure we consider is similar to the one employed 
in classical electrodynamics of continuous media in going from microscopic to 
macroscopic Maxwell equations. There are, however, important differences 
between these two procedures. In electrodynamics, each term of the microscopic 
Maxwell equations contains either derivative of electric or magnetic field, or 
charge/current density of the particles of the medium, but not the products 
of the two. This makes the averaging of the microscopic Maxwell equations 
technically simple. In contrast to this, there are terms in the microscopic 
neutrino evolution equation in medium that contain products of the neutrino 
wave function and the matter-induced potential of neutrinos (or the 
gradient of this potential). As the average of the product of two functions is 
in general different from the product of their averages, this is 
a nontrivial issue requiring special consideration.  

In the present paper we discuss the subtleties related to the averaging 
of the microscopic neutrino evolution equation in matter 
and establish the applicability domain of the standard macroscopic evolution 
equation. This, in particular, allows us to answer the question of when 
neutrino propagation in rarefied media (such as e.g.\ 
low-density gases or interstellar or intergalactic media) can be considered 
within the standard theory of neutrino flavour evolution in matter.

\section{\label{sec:micro} Microscopic neutrino evolution equation in matter}

Let us first consider the case of Dirac neutrinos; generalization to 
the Majorana neutrino case will be discussed in section~\ref{sec:Major}. 
The effective Lagrangian for neutrinos propagating in matter can be written 
as \cite{Halprin} 
\footnote{We use the natural units $\hbar=c=1$.} 
\be
{\cal L}=\bar{\nu}_L [i\slashed{\partial}-\slashed{\,V}(x)]\nu_L+ 
\bar{\nu}_R i\slashed{\partial}\nu_R-\bar{\nu}_LM\nu_R
-\bar{\nu}_R M^\dag\nu_L\,,
\label{eq:L1a}
\ee
where $\nu_L$ and $\nu_R$ are the left-handed (LH) and right-handed (RH) 
neutrino fields, $V^\mu(x)$ is the 4-vector potential 
induced by coherent neutrino forward scattering on matter constituents and 
$M$ is the neutrino mass matrix. Note that 
$V^\mu(x)$ and $M$ are matrices in flavour space, 
whereas $\nu_L$ and $\nu_R$ are flavour vectors. 
For definiteness, we shall speak about neutrino forward scattering on 
electrons, but our results will also 
apply to neutrino scattering on other matter constituents.   
The equations of motion for $\nu_L$ and $\nu_R$ 
following from the Lagrangian (\ref{eq:L1a}) are 
\begin{align}
& i\slashed{\partial}\nu_L-M\nu_R=\slashed{\,V}\nu_L\,, 
\label{eq:evol1L}
\\
& i\slashed{\partial}\nu_R-M^\dag\nu_L=0\,.
\label{eq:evol1R}
\end{align}
It will be convenient for us to use the chiral (Weyl) representation for the 
$\gamma$-matrices, in which $\gamma_5$ is diagonal and 
\be
\nu_L=\left(\begin{array}{c}
\phi \\ 0
\end{array}\right),\qquad
\nu_R=\left(\begin{array}{c}
0 \\ \chi
\end{array}\right),
\label{eq:notat1}
\ee
with $\phi$ and $\chi$ being the 2-component 
LH and RH neutrino fields, respectively. 
Eqs.~(\ref{eq:evol1L}) and (\ref{eq:evol1R}) then can be written as 
\begin{align}
& (i\partial^0-i\vec{\sigma}\vec{\nabla})\phi-M\chi=
(V^0+\vec{\sigma}\vec{V})\phi\,,
\label{eq:evol3L} \\
& (i\partial^0+i\vec{\sigma}\vec{\nabla})\chi-M^\dag\phi=0\,.
\label{eq:evol3R}
\end{align} 
We shall be assuming that the 4-vector of matter-induced potentials $V^\mu(x)$ 
depends on coordinate but does not change with time:
\be
V^\mu(x)=V^\mu(\vec{x})\,.
\label{eq:V}
\ee 
One can then look for the solutions of eqs.~(\ref{eq:evol3L})
(\ref{eq:evol3R}) in the form 
$\phi(x)=e^{-iEt}\phi(\vec{x})$, $\chi(x)=e^{-iEt}\chi(\vec{x})$.%
\footnote{These solutions can then be used as a basis for constructing 
neutrino wave packets with nonvanishing energy spread.}  
Substituting this into eqs.~(\ref{eq:evol3L}) and (\ref{eq:evol3R}), we 
obtain  
\begin{align}
& (E-i\vec{\sigma}\vec{\nabla})\phi-M\chi=
(V^0+\vec{\sigma}\vec{V})\phi\,,
\label{eq:evol4L} \\
& (E+i\vec{\sigma}\vec{\nabla})\chi-M^\dag\phi=0\,.
\label{eq:evol4R}
\end{align}
We shall be assuming that neutrinos are relativistic with $E\gg M,|V^\mu|$. 
Eliminating the RH neutrino field $\chi$ from eqs.~(\ref{eq:evol4L}) and 
(\ref{eq:evol4R}), to lowest order in  $V^\mu/E$ and 
$MM^\dag/E^2$ we \vspace*{-1.2mm}find 
\be
\big[\vec{\nabla}^2+E^2-MM^\dag-2E(V^0-\vec{v}\vec{V})
-i\vec{\sigma}[\vec{\nabla}(V^0+\vec{\sigma}\vec{V})]\big]\phi=0\,,
\label{eq:evol6a}
\ee  
where $\vec{v}$ is the neutrino velocity (see Appendix~{\ref{sec:AppA} for a 
detailed derivation). Note that, although this equation was derived for the 
neutrino field, the neutrino wave function $\langle 0|\phi(\vec{x})|\nu\rangle$ 
satisfies the same evolution equation. In what follows we shall 
consider the flavour 
evolution of the neutrino wave function, for which for conciseness we will use 
the same notation $\phi(\vec{x})$ as was up to now used for the neutrino field. 

\newpage
The neutrino potential $V^\mu$ in general contains both polar-vector and 
axial-vector contributions: $V^\mu=V^{\mu V}+V^{\mu A}$ \cite{Wolf}. 
To simplify our discussion, we shall mainly concentrate on neutrino 
propagation in media consisting of nonrelativistic (or randomly moving) and 
unpolarized particles. This is the case in many applications of interest, such 
as oscillations of solar, atmospheric or accelerator neutrinos. The spatial 
component $\vec{V}$ of the neutrino potential  
as well as the axial-vector part $V^{0A}$ of its time component 
can then be neglected. The case of non-vanishing $\vec{V}$ and $V^{0A}$} 
will be briefly discussed in section~\ref{sec:spatial}. 

With $\vec{V}$ set equal to zero, eq.~(\ref{eq:evol6a}) becomes  
\be
\big[\vec{\nabla}^2+E^2-MM^\dag-2EV^0-i(\vec{\sigma}\vec{\nabla}V^0)\big]
\phi=0\,.
\label{eq:evol6c}
\ee  
The standard approach is then to assume that the potential $V^0$ varies very 
slowly on the length scales of the order of the neutrino de Broglie wavelength 
$\sim 1/E$ and neglect the term containing the gradient of the potential. 
Eq.~(\ref{eq:evol6a}) then takes the form 
\be
\big[F+\vec{\nabla}^2\big]\phi=0\,, 
\label{eq:evol7}
\ee 
where 
\be
F = E^2-MM^\dag-2EV^0\,.
\label{eq:F1}
\ee
For one-dimensional neutrino motion along the $z$-axis, eq.~(\ref{eq:evol7}) 
can be factorized as \cite{Halprin} 
\be
\Big[\sqrt{F}-i\frac{d}{dz}\Big]\Big[\sqrt{F}+i\frac{d}{dz}\Big]\phi(z)=0\,,
\label{eq:evol8}
\ee
where, to lowest order in $MM^\dag/E^2$ and $V^0/E$, 
\be
\sqrt{F}=E-\frac{MM^\dag}{2E}-V^0(z)\,.
\label{eq:F2}
\ee
Note that the term $E$ here does not affect neutrino flavour transitions 
and can be omitted. For neutrinos propagating in the positive direction of the 
$z$-axis, eq.~(\ref{eq:evol8}) reduces to 
\be
\Big[i\frac{d}{dz}+\sqrt{F}\Big]\phi(z)=0\,.
\label{eq:evol9}
\ee
This equation, with $\sqrt{F}$ from eq.~(\ref{eq:F2}), coincides in form with 
the standard evolution equation for neutrino oscillations in matter, the 
difference being that eq.~(\ref{eq:evol9}) is actually a microscopic equation, 
while the neutrino evolution equation is usually interpreted as the 
macroscopic one. 

In refs.~\cite{cardchung,AW} neutrino flavour evolution in nonuniform matter 
was studied by considering the in-medium neutrino propagator. 
No assumption of one-dimensional neutrino propagation was explicitly made; 
however, it was assumed that the condition $E|\vec{x}|\gg 1$ is satisfied, 
that is, the distance $|\vec{x}|$ from the neutrino production point is large 
compared to the neutrino de Broglie wavelength. Under this condition 
the evolution is in fact one-dimensional. It was also assumed that the terms 
containing the gradient of the neutrino potential can be neglected.  
The obtained evolution equation coincides with eq.~(\ref{eq:evol9}) 
with the potential $V^0$ in the expression for 
$\sqrt{F}$ taken at the point $\vec{x}$ and $d/dz$ being the directional 
derivative along $\vec{x}$: $d/dz\equiv\vec{n}_{\vec{x}}\cdot\!\vec{\nabla}$, 
where $\vec{n}_{\vec{x}}$ 
is the unit vector in the \mbox{direction of $\vec{x}$.}

\section{\label{sec:macro}Averaging procedure and
macroscopic evolution equation }

In the above derivation of the neutrino evolution equation in matter 
we had to assume the potential $V^0(\vec{x})$ to be a slowly varying 
function of coordinate on the length scale of the order of the neutrino 
de Broglie wavelength $\lambda_D\sim 1/E$. As was discussed in the 
Introduction, this is in general not justified for microscopic  
neutrino potentials. We will therefore consider now the transition from 
microscopic to macroscopic description of neutrino flavour evolution 
in matter by averaging the microscopic evolution equation (\ref{eq:evol6c}). 
To this end, we will integrate it over a small but macroscopic volume $v_0$ 
around each point 
$\vec{x}$. The volume $v_0$ should be sufficiently large to contain 
a large number of particles of the medium, but small enough 
such that the macroscopic characteristics of the medium be nearly constant 
within it.%
\footnote{
\label{fnote:v0}
By this we mean that if we divide $v_0$ arbitrarily into two or more 
sub-volumes that are still macroscopic, the intensive macroscopic 
characteristics of the medium (such as 
density or temperature) in each of them will to a high accuracy be the same.}
We will be using the ``hat" notation for the coarse-grained (averaged) 
quantities, that is, for any integrable in $v_0$ function $g(\vec{x})$ 
we define
\be
\hat{g}(\vec{x})\,\equiv\,\frac{1}{v_0}\int_{v_0}d^3 x' g(\vec{x}+\vec{x}\,')\,.
\label{eq:aver1}
\ee
Obviously, differentiation and averaging 
operations commute, that is, the average of a derivative is equal to the 
derivative of the average. In particular, 
\be
\widehat{(\vec{\nabla}g)}(\vec{x})=\vec{\nabla}\hat{g}(\vec{x})\,.
\label{eq:aver2}
\ee
Thus, the averaging of all terms in eq.~(\ref{eq:evol6c}) except those 
containing the products of the neutrino 
wave function $\phi$ and the potential $V^0$ 
or its gradient is straightforward. In particular, upon the averaging,
\be
\big(\vec{\nabla}^2+E^2-MM^\dag\big)\phi(\vec{x}) \quad\longrightarrow \quad
\big(\vec{\nabla}^2+E^2-MM^\dag\big)\hat{\phi}(\vec{x})\,. 
\label{eq:aver3}
\ee

Before turning to the averaging of the remaining terms in 
eq.~(\ref{eq:evol6c}), let us consider the microscopic potential 
$V^0(\vec{x})$ and its averaging. We shall assume here the electrons of 
the medium to be pointlike particles with coordinates $\vec{x}_i$; the case 
when the electrons are described by atomic wave functions is discussed in 
Appendix~\ref{sec:AppB}. 

For definiteness, we consider 
the potential due to coherent forward neutrino-electron scattering 
mediated by  
weak charged currents, though the exact 
nature of the underlying interaction is not important for our discussion. 
The microscopic neutrino potential is then 
\be
V^0(\vec{x})=\sqrt{2}G_F v_F \sum_i\delta^3(\vec{x}-\vec{x}_i)\,,
\vspace*{-1.0mm}
\label{eq:V1}
\ee
where $G_F$ is the Fermi constant and $v_F$ is a coordinate-independent matrix 
characterizing the flavour structure of $V^0(\vec{x})$. In the flavour 
basis $(\nu_e, \nu_\mu, \nu_\tau)$ we have  $v_F\equiv {\rm diag}(1,\,0,\,0)$. 
Let the total number of electrons inside the averaging volume $v_0$ around 
the point $\vec{x}$\, be $N_0(\vec{x},v_0)$. As the volume $v_0$ is chosen 
to be sufficiently small to ensure that the matter density in it is 
essentially constant, $N_0(\vec{x},v_0)$ is proportional to $v_0$, i.e.\ the 
ratio 
\be
n_e(\vec{x})
~\equiv~\frac{N_0(\vec{x},v_0)}{v_0}
\label{eq:Ne}
\ee
is $v_0$-independent. 
The quantity $n_e(\vec{x})$ is the macroscopic electron number density in the 
medium, which is a smooth 
function of coordinate. 
{}From eqs.~(\ref{eq:V1}) and ~(\ref{eq:aver1}), for the averaged potential 
$\hat{V}^0(\vec{x})$ we then find 
\be
\hat{V}^0(\vec{x})=\sqrt{2}G_F v_F \frac{1}{v_0}
\int_{v_0} d^3 x'
\sum_{i=1}^{N_0(\vec{x},v_0)}
\delta^3(\vec{x}+\vec{x}\,'-\vec{x}_i)
\,.
\label{eq:V2}
\ee
Here the sum is over all the electrons in $v_0$. 
The integration is trivial, and we obtain 
\be
\hat{V}^0(\vec{x})
=\sqrt{2}G_F v_F n_e(\vec{x})\,. 
\label{eq:V2a}
\ee
This is the standard Wolfenstein potential employed in most studies of 
neutrino flavour transformations in matter. 

We are now in a position to perform the averaging of 
the terms in eq.~(\ref{eq:evol6c}) that contain the potential $V^0(\vec{x})$ 
and its gradient. Consider first the term $2EV^0\phi(\vec{x})$. 
By definition 
\be
\widehat{(V^0 \phi)}(\vec{x})
=\frac{1}{v_0}\int_{v_0}d^3 x'\, 
V^0(\vec{x}+\vec{x}\,')\phi(\vec{x}+\vec{x}\,')\,.
\label{eq:hat1}
\ee
Substituting here $V^0(\vec{x})$ from eq.~(\ref{eq:V1}) yields 
\be
\widehat{(V^0 \phi)}(\vec{x})=
\hat{V}^0(\vec{x})[\hat{\phi}(\vec{x})]_{\rm M.C.}\,, 
\label{eq:hat1a}
\ee
where we have used eq.~(\ref{eq:V2a}) and denoted
\be
[\hat{\phi}(\vec{x})]_{\rm M.C.}\equiv
\frac{1}{N_0(\vec{x},v_0)}
\sum_{i=1}^{N_0(\vec{x},v_0)}
\phi(\vec{x}_i)\,.
\label{eq:MC1}
\ee
We shall be assuming that the electrons of the medium are randomly distributed 
in the volume $v_0$, i.e.\ that $\vec{r}_i$ are random coordinates in $v_0$ 
with the uniform probability distribution function.   

\vspace*{2.0mm}
\noindent
{\colorbox{lightgray}
{\parbox[t]{14.8cm}
{\em 
The quantity $[\hat{\phi}(\vec{x})]_{\rm M.C.}$ 
\!is then  
nothing but the basic Monte Carlo estimator for the integral 
defining 
$\hat{\phi}(\vec{x})$ \mbox{according to 
eq.~(\ref{eq:aver1})} \cite{mc}.}
}}

\vspace*{2.8mm}
\noindent
Generally, Monte Carlo integration of a function $g(\vec{x})$ gives 
an accurate result when the number $N_0$ of the points $\vec{x}_i$ at which 
$g(\vec{x})$ is sampled is large, with the relative integration error being 
${\cal O}(N_0^{-1/2})$ (see section~\ref{sec:MC} for a more detailed 
discussion). As the averaging volume $v_0$ contains a macroscopically large 
number of electrons, in all situations of practical interest $N_0$ is very 
large and one can safely neglect the difference between 
$[\hat{\phi}(\vec{x})]_{\rm M.C.}$ and $\hat{\phi}(\vec{x})$. 
Eq.~(\ref{eq:hat1a}) then becomes 
\be
\widehat{(V^0 \phi)}(\vec{x})=
\hat{V}^0(\vec{x})\hat{\phi}(\vec{x})\,.
\label{eq:hat2}
\ee
That is, even though the average of the product of $V^0(\vec{x})$ and 
$\phi(\vec{x})$ does not in general factorize into the product of their 
averages, such a factorization does take place with high accuracy  
under the conditions that the electrons of the medium are 
pointlike and are randomly distributed in the averaging volumes, and that 
the total number of electrons in each averaging volume $v_0$ is 
sufficiently large. In Appendix~\ref{sec:AppB} we show that the 
assumption of pointlike electrons can actually be lifted. 

Next, we consider the averaging of the term 
$[\vec{\sigma}\vec{\nabla}V^0(\vec{x})]\phi(\vec{x})$ in 
(\ref{eq:evol6c}). We have  
\begin{align}
&\widehat{[(
\vec{\nabla}V^0)\phi]}(\vec{x})
\equiv\frac{1}{v_0}
\int_{v_0}d^3 x'\, 
\big[\vec{\nabla}_{\vec{x}+\vec{x}'}V^0(\vec{x}+\vec{x}\,')\big]
\phi(\vec{x}+\vec{x}\,')\nonumber \\
&
=\frac{1}{v_0}
\left\{\int_{v_0}d^3 x'\, 
\vec{\nabla}_{\vec{x}+\vec{x}'}\big[V^0(\vec{x}+\vec{x}\,')
\phi(\vec{x}+\vec{x}\,')\big]-
\int_{v_0}d^3 x'\, 
V^0(\vec{x}+\vec{x}\,')\vec{\nabla}_{\vec{x}+\vec{x}'}
\phi(\vec{x}+\vec{x}\,')\right\}.
\label{eq:hat3}
\end{align}
Let us consider the two terms in the second line of eq.~(\ref{eq:hat3}). 
For the first term we have 
\be
\frac{1}{v_0}
\int_{v_0}d^3 x'\, 
\vec{\nabla}_{\vec{x}+\vec{x}'}[V^0(\vec{x}+\vec{x}\,')
\phi(\vec{x}+\vec{x}\,')]=
\vec{\nabla}\frac{1}{v_0}
\int_{v_0}d^3 x'\, 
V^0(\vec{x}+\vec{x}\,')\phi(\vec{x}+\vec{x}\,')=
\vec{\nabla}\big[\hat{V}^0(\vec{x})\hat{\phi}(\vec{x})\big]\,,
\label{eq:hat3a}
\ee
where in the last equality we have used eq.~(\ref{eq:hat2}).
Consider now the last term in the second line in eq.~(\ref{eq:hat3}). 
Repeating the arguments that led to eq.~(\ref{eq:hat2}), we find 
\be
\frac{1}{v_0}
\int_{v_0}d^3 x'\, 
V^0(\vec{x}+\vec{x}\,')\vec{\nabla}_{\vec{x}+\vec{x}'}
\phi(\vec{x}+\vec{x}\,')=
\hat{V}^0(\vec{x})\widehat{\vec{\nabla}\phi}(\vec{x})=
\hat{V}^0(\vec{x})\vec{\nabla}\hat{\phi}(\vec{x})\,.
\label{eq:hat3aa}
\ee
Finally, using eqs.~(\ref{eq:hat3a}) and (\ref{eq:hat3aa}) in 
(\ref{eq:hat3}) we obtain 
\be
\widehat{[(
\vec{\nabla}V^0)\phi]}(\vec{x})
=\vec{\nabla}\big[\hat{V}^0(\vec{x})\hat{\phi}(\vec{x})\big]-
\hat{V}^0(\vec{x})\vec{\nabla}\hat{\phi}(\vec{x})
=\big[\vec{\nabla}\hat{V}^0(\vec{x})\big]\hat{\phi}(\vec{x})\,. 
\label{eq:hat4}
\ee

We now have all the ingredients in order to perform the averaging of the 
microscopic equation (\ref{eq:evol6c}). Combining eqs.~(\ref{eq:aver3}), 
(\ref{eq:hat2}) and (\ref{eq:hat4})
yields 
\be
\big[\vec{\nabla}^2+E^2-2E\hat{V}^0-MM^\dag-
i(\vec{\sigma}\vec{\nabla}\hat{V}^0)\big]
\hat{\phi}(\vec{x}) 
=0\,. 
\label{eq:evol6d}
\ee  
As the coarse-grained potential 
$\hat{V}^0(\vec{x})$ is a continuous function of coordinate, one can now 
use the argument that the term containing the gradient of $\hat{V}^0(\vec{x})$ 
can be neglected if this potential changes slowly on the length scales of 
the order of the neutrino de Broglie wavelength. This condition is always 
satisfied in practice, and we therefore drop the last term in the square 
brackets in eq.~(\ref{eq:evol6d}). Following the same arguments that led from 
eq.~(\ref{eq:evol6c}) to ~(\ref{eq:evol9}) and dropping the irrelevant term 
$E$ from $\sqrt{\hat{F}}$, we finally arrive at 
\be
i\frac{d}{dz}\hat{\phi}(z)=\Big[\frac{MM^\dag}{2E}+\hat{V}^0(z)
\,\Big]\hat{\phi}(z)=0\,.
\label{eq:evol10}
\ee
This is the standard neutrino evolution equation that was, though without 
proper justification, used in most studies of neutrino flavour transformations 
in matter.

\section{\label{sec:MC}Accuracy of Monte Carlo integration} 

Let us now discuss the accuracy of approximating the integrals involved in 
the averaging procedure by their basic Monte Carlo estimators, such as the 
one defined in eq.~(\ref{eq:MC1}). We consider here 
the case of neutrino 
scattering on pointlike electrons 
studied in the previous section; 
the generalization to 
the case of 
neutrino scattering on electrons in atoms   
(or molecules) is given in Appendix~\ref{sec:AppB}.

In general, for random $\vec{x}_i$ uniformly distributed in $v_0$ 
the expected value of the quantity $[\hat{g}(\vec{x})]_{\rm M.C.}\equiv 
(1/{\cal N}_0)\sum_i^{{\cal N}_0}g(\vec{x}_i)$ 
coincides with $\hat{g}(\vec{x})$, and its variance scales as $1/{\cal N}_0$. 
The error of the Monte Carlo estimation of $\hat{g}(\vec{x})$ therefore 
scales as $({\cal N}_0)^{-1/2}$. The proof is very simple; we give it here for 
the particular instance of the coarse-grained neutrino wave function 
$\hat{\phi}(\vec{x})$ in the case of neutrino scattering on pointlike 
electrons. 

For the expected value of the quantity $[\hat{\phi}(\vec{x})]_{\rm M.C.}$ 
defined in eq.~(\ref{eq:MC1}) we have 
\be
{\rm E}\big([\hat{\phi}(\vec{x})]_{\rm M.C.}\big)=
\frac{1}{N_0(\vec{x},v_0)}\!\sum_{i=1}^{N_0(\vec{x},v_0)}\!
{\rm E}\big(\phi(\vec{x}_i)\big)=
\frac{1}{N_0(\vec{x},v_0)}\!\sum_{i=1}^{N_0(\vec{x},v_0)} \int_{v_0}d^3 x' 
\phi(\vec{x}+\vec{x}')
{\rm PDF}(\vec{x}')\,.
\label{eq:}
\ee
As the random variable $\vec{x}_i$ is uniformly 
distributed in $v_0$, its probability distribution function 
${\rm PDF}(\vec{x})=1/v_0$, and we obtain  
\be
{\rm E}\big([\hat{\phi}(\vec{x})]_{\rm M.C.}\big)
=\hat{\phi}(\vec{x})\,.
\ee 
That is, $[\hat{\phi}(\vec{x})]_{\rm M.C.}$ is an unbiased estimator of 
$\hat{\phi}(\vec{x})$.
The expression for the variance of  $[\hat{\phi}(\vec{x})]_{\rm M.C.}$ can be 
found similarly. Direct calculation yields 
\be
{\rm var}\big([\hat{\phi}(\vec{x})]_{\rm M.C.}\big)
=\frac{\sigma^2[\phi(\vec{x})]}{N_0(\vec{x},v_0)}\,,
\label{eq:var1}
\ee
where 
$\sigma^2[\phi(\vec{x})]$ is defined as 
\be
\sigma^2[\phi(\vec{x})]
\equiv \frac{1}{v_0}\int_{v_0}d^3x'|\phi(\vec{x}+\vec{x}')|^2-
\Big|\frac{1}{v_0}\int_{v_0}d^3x' \phi(\vec{x}+\vec{x}')\Big|^2
=\widehat{|\phi(\vec{x})|^2}-\big|\hat{\phi}(\vec{x})\big|^2.  
\label{eq:sigma}
\ee
This quantity characterizes the speed of variation
of $\phi(\vec{x})$ with coordinate 
in the averaging volume $v_0$. In 
particular, it vanishes for $\phi(\vec{x})=const$. 

Eq.~(\ref{eq:var1}) has a simple meaning. 
The statistical error (standard deviation) introduced when replacing the 
coarse-grained neutrino wave function $\hat{\phi}(\vec{x})$ by its basic Monte 
Carlo estimator (\ref{eq:MC1}) is 
$\sigma[\phi(\vec{x})]/\sqrt{N_0(\vec{x},v_0)}$. 
If the function $\phi(\vec{x})$ is nearly constant throughout the averaging 
volume $v_0$, the quantity $\sigma[\phi(\vec{x})]$ 
is strongly suppressed; 
in this case the Monte Carlo integration of $\phi(\vec{x})$ with even a 
single sampling point should give an accurate value of $\hat{\phi}(\vec{x})$. 
On the other hand, if $\phi(\vec{x})$ is a fast varying function in $v_0$, 
then 
$\sigma[\phi(\vec{x})]$ is 
of the order $|\hat{\phi}(\vec{x})|$ or even larger. In this case 
a large number $N_0$ of the sampling points is necessary to achieve a 
good accuracy of Monte Carlo integration. In the case we consider, 
the role of the sampling points is played by the coordinates of the 
electrons in the medium. As we assume the averaging volumes $v_0$ to be 
macroscopic, $N_0$ is typically 
$\gtrsim 10^{12}$,%
\footnote{We adopt the definition of macroscopic volumes as those of linear 
size $\gtrsim 1\,\mu$m (see, e.g., \cite{reif}, p.~2). Noting that the 
electron number density can be written as $n_e=N_A \rho Y_e\;{\rm cm}^{-3}$ 
where $N_A$ is the Avogadro constant, $\rho$ is the matter density in g/cm$^3$ 
and $Y_e$ is the number of electrons per nucleon, we find that for 
$\rho\sim 3$\,g/cm$^3$ and $Y_e\simeq 1/2$ the number of electrons 
in a volume \,$\sim(1\,\mu\rm{m})^3$ is of the order $10^{12}$.}  
and the approximation of replacing the averaging integrals by their basic 
Monte Carlo estimators is very accurate. A possible exception is the case of 
neutrino propagation in rarefied media, which will be discussed 
in the next section.

\section{\label{sec:rare}Neutrino flavour transitions in rarefied media }

The averaging volumes $v_0$ that we use in our coarse-graining procedure have 
to satisfy several requirements. On the one hand, 
to make a statistical description possible, $v_0$ must be large enough to 
contain macroscopically large numbers $N_0$ of the particles of the medium.  
Very large $N_0$ also allowed us to replace, in the course of the 
coarse-graining, some averaging integrals by their 
basic Monte Carlo estimators. And finally, this allowed us to drop the term 
proportional to $\vec{\nabla}\hat{V}^0(\vec{x})$ from the macroscopic neutrino 
evolution equation and reduce it to the standard form (\ref{eq:evol10}). 

On the other hand, $v_0$ must be small enough such that inside 
it one could consider the macroscopic characteristics of the medium (and, in 
particular, the number density of the particles) as nearly constant. There is, 
however, one more consideration that bounds $v_0$ from above. As 
detection processes do not allow exact determination of the coordinate of 
each neutrino detection event, 
the experiments yield the detection 
data averaged over the active volume of the detector or, in case the detector 
allows some position resolution, 
over the volume $v_d$ of the corresponding detection region. 
The experiments therefore probe 
the flavour content of the incoming neutrino state with the same spatial 
resolution. The volume $v_0$ used 
in the averaging of the microscopic neutrino evolution equation 
must not exceed the volume of the detection region $v_d$, as otherwise the 
coarse-graining procedure would be too rough 
to allow an accurate prediction of the outcome of the experiment. 

Let us discuss the consequences of this constraint. 
As before, we for definiteness consider 
the effects of coherent 
neutrino forward scattering on the electrons of the medium. Let $n_e$ be a 
characteristic electron number density in the medium that affects 
the flavour transformations of neutrinos in the course of their propagation. 
Requiring that $N_0=v_0 n_e$ be, say, 
of the order $10^{12}$ or larger, from $v_d> v_0$ we find 
$v_d n_e \gtrsim 10^{12}$. 
For the electron number density in the medium we therefore obtain the lower 
limit 
\be
n_e \gtrsim \frac{10^{12}}{v_d}\,. 
\label{eq:condit1}
\ee
The linear sizes of neutrino detectors are typically in the 
$\sim 1$ meter to 1\,km range, but the position resolution for the neutrino 
events is usually much better.  
Taking as an example $v_d\sim 1\,{\rm m}^3$, from eq.~(\ref{eq:condit1}) 
we find the lower limit on the electron number density   
$n_e \gtrsim 10^{6}\;{\rm cm}^{-3}$. If, however, the coordinate of the 
neutrino detection point is known with a cm accuracy, we find  
$n_e \gtrsim 10^{12}\;{\rm cm}^{-3}$. 
At the same time, the nuclear emulsion film technology allows 
coordinate resolution at a 
$\mu$m level \cite{opera}; in this case 
eq.~(\ref{eq:condit1}) 
yields the condition $n_e\gtrsim 10^{24}$\,cm$^{-3}$. 
For comparison, the electron number density in dry air at sea level at 
20$^\circ$\,C is $\sim 3.6\times 10^{20}\;{\rm cm}^{-3}$. Thus, 
one might conclude that neutrino oscillations in air may 
be considered within the standard approach based 
on the macroscopic evolution equation (\ref{eq:evol10}) provided that the 
position resolution of the detector is not better than $~\sim 15\,\mu$m.   

It is natural to ask, however, whether taking matter effects into account for 
neutrinos propagating in air (or in any other low-density medium) makes any 
sense at all. There are two issues to be examined. First, 
matter effects on neutrino oscillations are typically important when 
the Wolfenstein potential $\hat{V}^0$ is at least of the same order as the 
neutrino kinetic energy difference 
$\Delta m^2/(2E)$.%
\footnote{
A possible exception is the parametric resonance of neutrino oscillations in 
matter, see below.}
This means that low-density media are expected to affect flavour transitions 
of neutrinos of sufficiently high energy. The potential $\hat{V}^0$ can be 
written in convenient units as 
\be
\hat{V}^0=
{\sqrt{2}G_F n_e}\simeq1.267\times 10^{-37}
\Big(\frac{n_e}{\rm cm^{-3}}\Big)\,{\rm eV}\,.
\label{eq:V01}
\ee
Taking for the estimate $\Delta m^2$ to be the ``solar'' mass squared 
difference 
$(\simeq 7.5\times 10^{-5}\,{\rm eV}^2$), we find that in air the condition 
$\hat{V}^0\gtrsim \Delta m^2/(2E)$ is satisfied for neutrinos of energies 
$E\gtrsim 800$ GeV. A small fraction of atmospheric neutrinos as well as 
high-energy neutrinos of astrophysical origin satisfy this condition.  

However, for matter effects in neutrino oscillations 
to be noticeable, yet another condition has to be met:\ neutrinos must 
propagate sufficiently large distances $l$ in matter 
\cite{LuSm1,LuSm2}. 
In practical terms, this so-called ``minimal length condition" 
implies that $l$ must at least exceed the refraction length $l_0$ defined as  
\be
l_0\equiv\frac{2\pi}{\sqrt{2}G_F n_e}\simeq 
9.8\times 10^{32}
\Big(\frac{{\rm cm^{-3}}}{n_e}\Big)\, {\rm cm}\,.
\label{eq:refr}
\ee
For air we have $l_0\sim 3\times 10^7$ km, and so the 
effects of the earth's atmosphere on neutrino oscillations can obviously be 
neglected. Note that the minimum length condition $l\gtrsim l_0$ is quite 
universal; in particular, it has to be also satisfied 
for the parametric enhancement of neutrino oscillations, for which 
the matter-induced neutrino potential $\hat{V}^0$ may be much smaller than 
$\Delta m^2/(2E)$ \cite{param1,param2,param3,LuSm1}. 

What about other low-density media? Consider, e.g., astrophysical neutrinos 
propagating in outer space. 
In the interstellar medium, the average electron number density is about 
1\,cm$^{-3}$. 
As follows from eq.~(\ref{eq:refr}), the minimum length 
condition then implies that for the effects of the medium on neutrino flavour 
transitions to be noticeable, the neutrinos should propagate at least 
distances 
$l\sim 10^{33}$\,cm, which is about four orders of magnitude larger than the 
diameter of the observable Universe. Clearly, the effects of the 
interstellar medium on flavour evolution of astrophysical neutrinos can be 
safely neglected. 
More detailed discussion of the consequences of the minimum length condition 
on neutrino oscillations in various media can be found in \cite{LuSm1,LuSm2}.  

Thus, although in some rarefied media the number density of particles may be 
too low to allow the transition from microscopic to macroscopic description of 
neutrino flavour evolution, quite often the minimum length condition is then 
also violated. In those cases one can simply neglect all matter effects  
and consider neutrino oscillations as occurring in vacuum. 
Obviously, no averaging is needed in such situations.   

Let us now return to the lower bound~(\ref{eq:condit1}) on the electron number 
density that follows from the coarse-graining conditions and is related to the 
coordinate resolution of the detector. When is it more restrictive than the 
constraint coming from the minimum length condition $l\gtrsim l_0$ 
and so has to be taken into account? 
Comparing 
eqs.~(\ref{eq:condit1}) and~(\ref{eq:refr}), we find that this happens when  
\be
v_d^{1/3}\lesssim 10^{-7}\left(\frac{l}{1\,{\rm cm}}\right)^{1/3}\,{\rm cm}.
\label{eq:resol1}
\ee 
As an example, for the coordinate resolution of the neutrino detection  
$v_d^{1/3}\sim 1\mu$m the condition in eq.~(\ref{eq:condit1}) is more 
restrictive than 
the minimum length condition if the distance $l$ traveled by neutrinos in 
matter exceeds about $10^4$ km, which is of the 
same order as the diameter of the earth.  Eq.~(\ref{eq:condit1}) then requires 
$n_e\gtrsim 10^{24}\,{\rm cm}^{-3}$, that is, the average matter density 
should satisfy $\rho\gtrsim 3$\,g/cm$^3$, which is fulfilled for the 
matter of the earth.  

\section{\label{sec:spatial}
Media with bulk currents and magnetization} 
As was pointed out in section~\ref{sec:micro}, the spatial component $\vec{V}$ 
of the matter-induced neutrino potential and the axial-vector part 
of its time component, $V^{0A}$, have negligible effects on neutrino flavour 
transitions in media consisting of nonrelativistic or randomly moving 
particles with no spin polarization. 
They may, however, play an important role for neutrinos propagating in 
magnetized backgrounds or in media with bulk particle currents which may 
exist, e.g., at certain stages of supernova explosions. 
In this section we briefly discuss the averaging of the neutrino evolution 
equation with the $\vec{V}$ and $V^{0A}$ contributions included.

Consider first neutrino propagation in a medium with unpolarized electrons; 
the axial-vector part of the neutrino potential then vanishes. We will assume, 
however, that there are macroscopic electron currents in the medium.  
In this case one should take into account the spatial component of the 
polar-vector part of the neutrino potential. For pointlike electrons we have 
\be
\vec{V}^V=
\vec{V}^V(\vec{x},\{\vec{v}_i\})=
\sqrt{2}G_F v_F \sum_i\vec{v}_i\delta^3(\vec{x}-\vec{x}_i)\,,
\vspace*{-1.0mm}
\label{eq:VV1}
\ee
where $\vec{v}_i$ is the velocity of the $i$th electron. For 
velocity-dependent quantities the averaging procedure of eq.~(\ref{eq:aver1}) 
has to be modified: in addition to the spatial averaging, it should include 
the averaging over the electron velocities. Keeping the same ``hat'' notation 
for the averaged quantities as before, we now define the average of a function 
$g(\vec{x},\vec{v})$ as  
\be
\hat{g}(\vec{x})
\,\equiv\,\frac{1}{v_0}\int_{v_0}d^3 x' 
\int d^3v\, f_{\vec{x}}(\vec{v})g(\vec{x}+\vec{x}'\,,\vec{v})\,.
\label{eq:aver1a}
\ee
Here $f_{\vec{x}}(\vec{v})$ is the electron velocity distribution function, 
which we assume to be time-independent%
\footnote{Actually, it will be sufficient for us to assume that the velocity 
distribution changes negligibly over the time intervals of the order of the 
time of neutrino passage through the averaging volume $v_0$.}
and normalized according to $\int d^3v f_{\vec{x}}(\vec{v})=1$. 
Note that it may be different in different parts of the system but, as any 
other intensive macroscopic characteristic of the medium, it is essentially 
position-independent over distances of the order of the linear size of $v_0$.%
\footnote{The velocity distribution function $f_{\vec{x}}(\vec{v})$ can 
be considered as an average of the phase space density ${\cal F}
(\vec{x},\vec{v})$ over spatial volumes of the order of $v_0$ \cite{klim}.}
For this reason we use $f_{\vec{x}}(\vec{v})$ rather than 
$f_{\vec{x}+\vec{x}\,'}(\vec{v})$ in the integrand in eq.~(\ref{eq:aver1a}).
The averaged quantities, however, in general depend on the position of the 
point $\vec{x}$ on which the volume $v_0$ is centered. In particular, for the 
average of the microscopic potential~(\ref{eq:VV1}) we find 
\be
\hat{\vec{V}}^V(\vec{x})=\sqrt{2}G_F v_F 
n_e(\vec{x})\hat{\vec{v}}_e(\vec{x})\,,
\label{eq:hatV}
\ee 
where 
\be
\hat{\vec{v}}_e(\vec{x})=\int d^3v\, f_{\vec{x}}(\vec{v})\vec{v}\,
\label{eq:hatv}
\ee 
is the macroscopic local electron velocity. 

Let us now discuss the averaging of the evolution equation (\ref{eq:evol6a}). 
Obviously, for velocity-independent terms of this equation the averaging 
procedure (\ref{eq:aver1a}) gives the same results as the one in 
eq.~(\ref{eq:aver1}). Consider the averaging of the terms proportional to 
$\vec{V}^V\phi$. Straightforward calculation gives 
\be
\widehat{(\vec{V}^V \phi)}(\vec{x})
=\hat{\vec{V}}^V(\vec{x})[\hat{\phi}(\vec{x})]_{\rm M.C.}\, 
\label{eq:Vvechat1}
\ee
with $[\hat{\phi}(\vec{x})]_{\rm M.C.}$ defined in eq.~(\ref{eq:MC1}).  
As was discussed in section~\ref{sec:macro}, since the averaging volume $v_0$ 
contains a macroscopically large number of randomly distributed electrons,  
one can safely replace $[\hat{\phi}(\vec{x})]_{\rm M.C.}$ by 
$\hat{\phi}(\vec{x})$, and eq.~(\ref{eq:Vvechat1}) becomes  
$\widehat{(\vec{V}^V \phi)}(\vec{x})
=\hat{\vec{V}}^V(\vec{x})\hat{\phi}(\vec{x})$. 
The averaging of the terms in eq.~(\ref{eq:evol6a}) containing the spatial 
derivatives of $\vec{V}^V$ is then done similarly to the averaging of 
$(\vec{\nabla}V^0)\phi$ performed in section~\ref{sec:macro}, and we obtain 
\be
\widehat{[(
\nabla^i \vec{V}^V)\phi]}(\vec{x})
=\big[\nabla^i \hat{\vec{V}}^V(\vec{x})\big]\hat{\phi}(\vec{x})\,. 
\label{eq:hat4a}
\ee

Let us now turn to neutrino propagation in magnetized media. In this case 
the electrons of the medium have non-vanishing average spin polarization, 
which leads to a non-zero axial-vector contribution to the macroscopic 
neutrino potential $\hat{V}^\mu$. The averaging procedure of 
eq.~(\ref{eq:aver1a}) will then have to be modified in the following way:
\begin{itemize}
\item
The electron velocity distribution function $f_{\vec{x}}(\vec{v})$ 
should be replaced by the electron velocity and spin distribution 
function $f_{\vec{x}}(\vec{v},\vec{s})$, where 
$\vec{s}=\psi_e^\dag\vec{\sigma}_e\psi_e$ is the electron polarization 
vector, $\psi_e$ being the 2-component electron spinor.
\item
In addition to integrations over $\vec{x}\,'$ and $\vec{v}$, the summation 
over the electron polarization states should be performed. 
\end{itemize}

\noindent
The averaging of the neutrino evolution equation (\ref{eq:evol6a}) is 
then done along the same lines as discussed above. 
The resulting macroscopic evolution equation has the same form as 
eq.~(\ref{eq:evol6a}), but with the $V^0$, $\vec{V}$ and $\phi$ replaced 
by $\hat{V}^0$, $\hat{\vec{V}}$ and $\hat{\phi}$ respectively. As discussed 
in section~\ref{sec:macro}, the terms with spatial derivatives of the 
components of the macroscopic neutrino potential $\hat{V}^\mu$ can then be 
neglected, and the evolution equation can be reduced to a first order one. 
It coincides in form with that in eq.~(\ref{eq:evol10}), except 
that the macroscopic potential $\hat{V}^0$ has to be replaced by 
$\hat{V}\equiv \hat{V}^0-\vec{v}\hat{\vec{V}}$, where the coarse-graining 
procedure now in general includes, in addition to the spatial averaging,  
the averaging over the velocities and spin polarizations of the background 
electrons.

\section{\label{sec:Major} The case of Majorana neutrinos}

In section~\ref{sec:micro}, the microscopic evolution equation describing 
neutrino oscillations in matter was derived 
in the Dirac neutrino case from the equations of motion 
for the LH and RH neutrino fields, 
(\ref{eq:evol4L}) and (\ref{eq:evol4R}). 
For Majorana neutrinos, the LH and RH neutrino fields are not independent: 
they are related by $\chi=-i\sigma_2\phi^*$. 
Equations of motion 
(\ref{eq:evol3L}) and (\ref{eq:evol3R}) 
then have to be replaced by 
\begin{align}
& (i\partial^0-i\vec{\sigma}\vec{\nabla})\phi-
M(-i\sigma_2\phi*)=
(V^0+\vec{\sigma}\vec{V})\phi\,,
\label{eq:evol3La} \\
& (i\partial^0+i\vec{\sigma}\vec{\nabla})
(-i\sigma_2\phi*)-M^*\phi=
(-V^0+\vec{\sigma}\vec{V})(-i\sigma_2\phi*)\,.
\label{eq:evol3Ra}
\end{align}
Note that eqs.~(\ref{eq:evol3La}) and (\ref{eq:evol3Ra}) are actually 
equivalent.\ 
Unlike its Dirac-case analogue~(\ref{eq:evol3R}),  
eq.~(\ref{eq:evol3Ra}) depends on the 
neutrino potentials 
$(V^0, \vec{V})$; 
this is because, in contrast to the Dirac neutrino case, 
for Majorana neutrinos the RH neutrino states are not sterile. 

As in section~\ref{sec:micro}, we look for the solution of the equations of 
motion in the form $\phi(x)=e^{-iEt}\phi(\vec{x})$. 
Eliminating $\phi^*$ from eqs.~(\ref{eq:evol3La}) and~(\ref{eq:evol3Ra}), 
to lowest order in  $V^0/E$ and $MM^\dag/E^2$ we again obtain 
eq.~(\ref{eq:evol6a}). Its averaging and the transition to the first-order 
neutrino evolution equation are then carried out exactly as in the Dirac 
neutrino case, leading to the same macroscopic evolution equation. Thus, 
under our assumption that neutrinos are relativistic with $E\gg M, |V^\mu|$, 
flavour transitions of Dirac and Majorana neutrinos in matter are 
described by the same evolution equation.

\section{\label{sec:disc}Summary and discussion}

In this paper we considered a transition from microscopic (fine-grained) 
to macroscopic (coarse-grained) description of neutrino flavour transitions 
in matter through a proper averaging in coordinate space. Our primary 
motivation was 
to justify neglecting the term proportional to the gradient of the 
potential in the neutrino evolution equation, which 
allows one to reduce this equation to the standard form (\ref{eq:evol10}). 
However, the transition to a statistical (macroscopic) description 
is also necessary 
on more general grounds: it is not possible in practice to solve the 
microscopic evolution equation, not even to mention that this would require 
the knowledge of the coordinates of all the matter constituents. 
As neutrino experiments 
yield the detection data averaged over macroscopic volumes 
determined by the coordinate resolution of the detectors, 
coarse-grained neutrino wave functions are 
adequate for the practical purposes of predicting the expected outcomes of 
the experiments or interpreting the obtained data. 

In all the previous studies of neutrino flavour transitions in matter it was 
implicitly assumed that some averaging of the microscopic neutrino potential 
has already been done. A consistent 
approach, however, would require to average the microscopic evolution equation 
as a whole and not just the neutrino potential.  
To the best of our knowledge, 
no such averaging has been carried out in the past. 

In the present paper we performed the averaging of the microscopic neutrino 
evolution equation by integrating it over a small but macroscopic volume $v_0$ 
around each point in coordinate space. The choice of the averaging volume was 
dictated by a number of factors. 
On the one hand, 
it must be large enough to contain macroscopically large numbers $N_0$ of 
the particles of the medium.  This allows a statistical 
description of the medium. 
On the other hand, $v_0$ must be small enough so that inside it one could 
consider the intensive macroscopic characteristics of the medium as nearly 
constant. Another upper limit on the averaging volume $v_0$ comes from 
experimental considerations: it should not exceed the volume $v_d$ of the 
detection region that is 
determined by the spatial resolution of the detector. 
The same consideration, together with the requirement that the 
number of particles in the volume $v_0$ be macroscopically large, puts a 
lower limit on the number density of the particles in the medium.

In the course of the averaging of the microscopic neutrino evolution equation 
in matter one encounters a difficulty  
related to the presence 
of the terms containing the products of the neutrino potential or its 
gradient and the neutrino wave function. 
As the average of the product of two functions is in general 
different from the product of their averages, 
such terms require special consideration. 
We have demonstrated that 
for the product terms in the neutrino evolution equation the factorization 
does take place with very high accuracy 
provided that the electrons of the medium are randomly distributed in the 
averaging volumes and that the total number $N_0$ of electrons in each 
averaging volume $v_0$ is macroscopically large. 
Our key observation was 
that under these conditions 
one can replace the integral over the averaging volume $v_0$ of the neutrino 
wave function by its basic Monte Carlo estimator, which immediately leads 
to the desired factorization.  

We have also established a lower bound on the number density of the particles 
of the medium that has to be satisfied in order for the coarse-graining 
procedure to be adequate to experiments with a given spatial resolution 
of the neutrino detection events. 
This bound, in principle, establishes under what conditions neutrino 
oscillations in low-density media can be described by the standard macroscopic 
evolution equation. This condition, however, becomes irrelevant if the matter 
density is so small that the distance neutrinos propagate in it is small  
compared to the refraction length $l_0$ defined in eq.~(\ref{eq:refr}). 
This is because in this case matter effects on neutrino oscillations can be 
safely neglected. 

To conclude, by performing a coarse-graining of the microscopic neutrino 
evolution equation in matter we have derived and justified the standard 
macroscopic evolution equation (\ref{eq:evol10}) which was previously used 
without proper justification. We have also found the validity conditions 
for this equation. In addition to the usual requirement that the neutrinos 
must be relativistic with $E\gg M, |V^\mu|$, we had to assume that the 
averaging volumes $v_0$ contain macroscopically large numbers of electrons 
which are distributed randomly within $v_0$. Our treatment is therefore 
not applicable to the case of neutrino propagation in media that are ordered 
at the 
subatomic level, such as crystals.%
\footnote{Note that for 
such media it is sometimes possible to solve microscopic neutrino 
evolution equations 
\cite{mt}.} 
It is, however, still valid for the 
macroscopically ordered media, such as e.g.\ periodic structures with 
macroscopic periods of density modulation, including those that lead to 
parametric enhancement of neutrino oscillations in matter. 

\acknowledgments 

The author thanks Alexei Smirnov for very useful 
discussions. 
 
\appendix
\section{\label{sec:AppA} Derivation of eq.~(\ref{eq:evol6a})}
Acting on eq.~(\ref{eq:evol4L}) with $(E+i\vec{\sigma}\vec{\nabla})$ and 
making use of eq.~(\ref{eq:evol4R}), we \vspace*{-1.2mm}obtain 
\be
\big[\vec{\nabla}^2+E^2-MM^\dag\big]\phi=
(E+i\vec{\sigma}\vec{\nabla})\big[(V^0+\vec{\sigma}\vec{V})\phi\big]\,.
\label{eq:evolA1}
\ee  
For the right-hand side of this equation we have 
\begin{align}
{\rm RHS}=&
E(V^0+\vec{\sigma}\vec{V})\phi+i(\vec{\sigma}\vec{\nabla}V^0)\phi
+V^0(i\vec{\sigma}\vec{\nabla}\phi)
+i\vec{\sigma}\vec{\nabla}(\vec{\sigma}\vec{V}\phi)\nonumber\\
=&E(V^0+\vec{\sigma}\vec{V})\phi+i(\vec{\sigma}\vec{\nabla}V^0)\phi
+V^0(i\vec{\sigma}\vec{\nabla}\phi)
+i[\vec{\sigma}\vec{\nabla}(\vec{\sigma}\vec{V})]\phi
+i\sigma^i\sigma^k V^k\nabla^i\phi \nonumber\\
=&E(V^0+\vec{\sigma}\vec{V})\phi+i[\vec{\sigma}
\vec{\nabla}(V^0+\vec{\sigma}\vec{V})]\phi
+V^0(i\vec{\sigma}\vec{\nabla}\phi) 
+i[-\sigma^k\sigma^i+2\delta^{ik}]V^k\nabla^i\phi\nonumber\\
=&E(V^0+\vec{\sigma}\vec{V})\phi+i[\vec{\sigma}
\vec{\nabla}(V^0+\vec{\sigma}\vec{V})]\phi
+(V^0-\vec{\sigma}\vec{V})(i\vec{\sigma}\vec{\nabla}\phi)
+2\vec{V}(i\vec{\nabla}\phi)
\,.
\label{eq:rhs1}
\end{align}
We assume neutrinos to be relativistic with $E\gg M,|V^\mu|$ 
and evaluate RHS to leading order in $MM^\dag/E^2$ and $|V^\mu|/E$. As follows 
from eqs.~(\ref{eq:evol4L}) and (\ref{eq:evol4R}), the expression 
$i\vec{\sigma}\vec{\nabla}\phi$, which enters in eq.~(\ref{eq:rhs1}) 
multiplied by the components of the neutrino potential $V^\mu$, can then be 
replaced there by $E\phi$. In addition, one can replace in eq.~(\ref{eq:rhs1}) 
the quantity $i\vec{\nabla}\phi$ by $-\vec{p}\phi=-E\vec{v}\phi$, where 
$\vec{p}$ is the neutrino momentum and $\vec{v}$ is its velocity.%
\footnote{Strictly speaking, propagating neutrinos 
are described not by plane waves of momentum $\vec{p}$, but by wave packets 
with mean momentum $\vec{p}$ and momentum uncertainty 
$\sigma_p$. Therefore, $i\vec{\nabla}\phi=[-\vec{p}+{\cal O}(\sigma_p)]\phi$. 
However, because $\sigma_p\ll |\vec{p\,}|$ and the expression 
$i\vec{\nabla}\phi$ enters in eq.~(\ref{eq:rhs1}) multiplied by $\vec{V}$,  
the term proportional to $\sigma_p$ is subleading and can 
to the adopted accuracy be neglected.}
Taking this into account, from the last line in eq.~(\ref{eq:rhs1}) we find  
\be
{\rm RHS}=
2E(V^0-\vec{v}\vec{V})\phi+i[\vec{\sigma}\vec{\nabla}(V^0+\vec{\sigma}
\vec{V})]\phi\,.
\label{eq:rhs2}
\ee
Substituting this for the right-hand side of eq.~(\ref{eq:evolA1}), we arrive 
at eq.~(\ref{eq:evol6a}). 

\section{\label{sec:AppB} Neutrino coherent forward scattering on electrons 
in atoms and molecules}

We generalize here the results of sections~\ref{sec:macro} and \ref{sec:MC} 
to the case of neutrino forward scattering on atomic and molecular electrons. 
The results of section~\ref{sec:spatial} can be generalized quite similarly. 

\subsection{\label{sec:pointAt} Approximation of pointlike atoms}

Consider first the idealized situation when one can neglect the size 
of the atoms and 
treat them as pointlike objects. This case is similar to the 
one considered in sections~\ref{sec:macro} and \ref{sec:MC}, the 
difference being that the 
scatterers may now have electron numbers different from one. 
 
Let the medium 
consist of $K$ types of 
pointlike objects (scatterers) with the electron numbers $Z_k$ 
($k=1,...,K$), and let the total numbers of the scatterers of 
the $k$th type in the medium be $N_k$. 
The microscopic neutrino potential is in this case 
\be
V^0(\vec{x})=\sqrt{2}G_F v_F
\sum_{k=1}^K Z_k 
\sum_{i=1}^{N_k}\delta^3(\vec{x}-\vec{x}_i)\,.
\label{eq:V3}
\ee
Its averaging 
leads to the standard Wolfenstein potential 
(\ref{eq:V2a})
with the macroscopic electron number density 
\be
n_e(\vec{x})=\frac{1}{v_0}\sum_{k=1}^K Z_k N_{0k}(\vec{x},v_0)\,.
\label{eq:Ne2}
\ee
Here $N_{0k}(\vec{x},v_0)$ is the number of the scatterers of the $k$th 
type in the averaging volume $v_0$ around the point $\vec{x}$, 
so that the sum in (\ref{eq:Ne2}) is just the total number of electrons in 
$v_0$. Substituting the expression for $V^0(\vec{x})$ from 
eq.~(\ref{eq:V3}) into eq.~(\ref{eq:hat1}) yields eq.~(\ref{eq:hat1a}),  
where $[\hat{\phi}(\vec{x})]_{\rm M.C.}$ is now given by 
\be
[\hat{\phi}(\vec{x})]_{\rm M.C.}\equiv
\frac{\sum_{k=1}^{K}Z_k\sum_{i=1}^{N_{0k}(\vec{x},v_0)}
\phi(\vec{x}_i)}{\sum_{k=1}^K Z_k N_{0k}(\vec{x},v_0)}\,.
\label{eq:MC2}
\ee
As before, we assume that all the 
scatterers are randomly and uniformly distributed in the averaging volumes 
$v_0$; the obtained results then essentially coincide with 
those of section~\ref{sec:macro} and \ref{sec:MC}. 
In particular, the macroscopic neutrino evolution equation 
is again given by eq.~(\ref{eq:evol10});  
it is also easy to show that the expected value of 
$[\hat{\phi}(\vec{x})]_{\rm M.C.}$ coincides with $\hat{\phi}(\vec{x})$. 
For the variance of $[\hat{\phi}(\vec{x})]_{\rm M.C.}$ we find 
\be
{\rm var}\big([\hat{\phi}(\vec{x})]_{\rm M.C.}\big)=
\frac{\sum_{k=1}^K Z_k^2 N_{0k}(\vec{x},v_0)}
{\big[\sum_{k=1}^K Z_k N_{0k}(\vec{x},v_0)\big]^2}
\sigma^2[\phi(\vec{x})]\,.
\label{eq:var2}
\ee
This quantity typically scales as $\sim 1/N_{0j}$,%
\footnote{\label{footexcept} Except when $Z_j^2N_{0j}\ll Z_k^2N_{0k}$ 
for some $k\ne j$; in that case 
${\rm var}\big([\hat{\phi}(\vec{x})]_{\rm M.C.}\big)$ scales as 
$(Z_k^2/Z_j^2)(N_{0k}/N_{0j}^2)$.}
where 
$N_{0j}$ is the number of the scatterers with the largest 
$Z_j N_{0j}$ contained in the averaging volume $v_0$. 

\subsection{\label{sec:atomsFinite} Neutrino forward scattering on 
atoms and molecules of finite size}

Let us now lift our assumption of pointlike atoms and consider a medium 
consisting of atoms of finite size described by atomic wave functions. 
As in the previous subsection, we shall consider a medium 
containing $K$ types of atoms with atomic numbers $Z_k$ ($k=1,...,K$). 
Let the atomic wave functions be 
$\Psi_k(\vec{x}_1,\dots \vec{x}_{Z_k};\,\vec{x}_0)$, where $\vec{x}_0$ 
is the coordinate of the center of the atom. 
For an atom with $Z_k$ electrons and the center at 
$\vec{x}_0$, the electron number density is 
\be
\rho_k(\vec{x}, \vec{x}_0)
=\sum_{a=1}^{Z_k}\int|\Psi_k(\vec{x}_1,\dots 
\vec{x}_{Z_k};\,\vec{x}_0)|^2\delta^3(\vec{x}-\vec{x}_a)
d^3 x_1\dots d^3x_{Z_k}\,. 
\label{eq:rho1}
\ee
We adopt the standard normalization convention in which 
the integrals of the squared moduli of the atomic wave functions are 
normalized to unity; 
$\rho_k(\vec{x}, \vec{x}_0)$ then satisfies  
\be
\int d^3 x \rho_k(\vec{x},\vec{x}_0)=Z_k\,.
\label{eq:norm}
\ee 
The microscopic neutrino potential is in this case 
\be
V^0(\vec{x})=\sqrt{2}G_F v_F
\sum_{k=1}^K 
\sum_{i=1}^{N_k} \rho_k(\vec{x},\,\vec{x}_i)\,.
\label{eq:V4}
\ee
The coarse-grained neutrino potential $\hat{V}^0(\vec{x})$ 
takes the standard form (\ref{eq:V2a}) with the macroscopic electron 
density $n_e(\vec{x})$ given by eq.~(\ref{eq:Ne2}), as in the case of neutrino 
forward scattering on pointlike atoms. The average of the product 
$V^0(\vec{x})\phi(\vec{x})$ has the same form as in eq.~(\ref{eq:hat1a}), but 
with $[\hat{\phi}(\vec{x})]_{\rm M.C.}$ defined as 
\be
[\hat{\phi}(\vec{x})]_{\rm M.C.}\equiv
\frac{\sum_{k=1}^{K}\sum_{i=1}^{N_{0k}(\vec{x},v_0)}
\Phi_k(\vec{x}, \vec{x}_i)}{\sum_{k=1}^K Z_k N_{0k}(\vec{x},v_0)}\,,
\label{eq:MC3}
\ee
where 
\be
\Phi_k(\vec{x}, \vec{x}_i)\equiv \int_{v_0} d^3 x'\,
\rho_k(\vec{x}+\vec{x}\,'\!, \vec{x}_i) \phi(\vec{x}+\vec{x}\,')\,.
\label{eq:Phi}
\ee 
It is easy to show that $[\hat{\phi}(\vec{x})]_{\rm M.C.}$ is an unbiased 
estimator of $\hat{\phi}(\vec{x})$, 
i.e.\ ${\rm E}\big([\hat{\phi}(\vec{x})]_{\rm M.C.}
\big)=\hat{\phi}(\vec{x})$. 
To prove this, let us first notice that due to translational invariance 
$\rho_k(\vec{x},\vec{x}_0)=\rho_k(\vec{x}-\vec{x}_0)$, so that in the 
normalization condition (\ref{eq:norm}) one can replace the integration over 
$\vec{x}$ by that over $\vec{x}_0$. Therefore,\\ 
\be
{\rm E}\big(\Phi_k(\vec{x}, \vec{x}_i)\big)=
\int d^3 x_i \Phi_k(\vec{x}, \vec{x}_i){\rm PDF}(\vec{x}_i) 
=Z_k \hat{\phi}(\vec{x})\,,
\label{eq:proof1}
\ee
where, as usual, we assumed that the scattering centers are randomly 
distributed in $v_0$ with the uniform probability distribution function.  
Together with the definition of $[\hat{\phi}(\vec{x})]_{\rm M.C.}$ in 
eq.~(\ref{eq:MC3}), this gives ${\rm E}\big([\hat{\phi}(\vec{x})]_{\rm M.C.}
\big)=\hat{\phi}(\vec{x})$. 
  
For the variance of $[\hat{\phi}(\vec{x})]_{\rm M.C.}$ we find 
\be
{\rm var}\big([\hat{\phi}(\vec{x})]_{\rm M.C.}\big)=
\frac{\sum_{k=1}^K Z_k^2 N_{0k}(\vec{x},v_0) 
\sigma^2[\Phi_k(\vec{x})]}
{\big[\sum_{k=1}^K Z_k N_{0k}(\vec{x},v_0)\big]^2}\,,
\label{eq:var3}
\ee
where 
\be
{\sigma^2}[\Phi_k(\vec{x})]
\equiv 
\frac{1}{Z_k^{2}v_0}
\int_{v_0}d^3x_i\left|\Phi_k(\vec{x},\vec{x}_i)\right|^2
\,-~|\hat{\phi}(\vec{x})|^2\,.
\label{eq:sigma2}
\ee
This quantity depends on the atomic wave function 
$\Psi_k(\vec{x}_1,\dots \vec{x}_{Z_k};\,\vec{x}_i)$ and, unlike 
$\sigma[\phi(\vec{x})]$,  
it cannot be expressed 
solely through $\phi(\vec{x})$. However, it shares with  
$\sigma[\phi(\vec{x})]$ the property of being strongly suppressed for 
nearly constant $\phi(\vec{x})$ and also vanishes when $\phi(\vec{x})$ 
is constant. Just like in the case of neutrino forward scattering on 
pointlike atoms, 
${\rm var}\big([\hat{\phi}(\vec{x})]_{\rm M.C.}\big)$ 
typically scales as $\sim 1/N_{0j}$, where 
$N_{0j}$ is the number of the scatterers with the largest 
$Z_j N_{0j}$ contained in the averaging volume $v_0$. 

As the averaging volumes $v_0$ are assumed to contain 
macroscopically large numbers of atoms, it is justified to replace the 
averaged quantities $\hat{\phi}(\vec{x})$ and 
$\vec{\nabla}\hat{\phi}(\vec{x})$ by their 
Monte Carlo estimators. Thus, for the case of 
neutrino forward scattering on atomic electrons the results of the 
averaging 
of the microscopic neutrino evolution equation 
coincide with those obtained in section~\ref{sec:macro}.
The difference is that for the Monte Carlo estimator of the macroscopic 
neutrino field $\hat{\phi}(\vec{x})$ 
we actually take, instead of a linear combination of the values of 
${\phi}$ at random coordinates $\vec{x}_i$ in $v_0$, a linear combination of 
the values of ${\phi}$ averaged over small (atomic size) volumes around the 
random coordinates of the centers of the atoms 
inside $v_0$, and similarly for $\vec{\nabla}{\phi}(\vec{x})$. 

We were assuming here that all the electrons of the medium are contained in 
atoms; it is easy to see, however, that our results are also directly 
applicable to the case of media consisting of arbitrary mixture of molecules, 
neutral atoms, ions and 
free electrons.

\end{document}